# An Optimization Method of Asymmetric Resonant Cavities for Unidirectional Emission

Fang-Jie Shu, Chang-Ling Zou, Fang-Wen Sun

*Abstract*—In this paper, we studied the repeatability and accuracy of the ray simulation for one kind of Asymmetric Resonant Cavities (ARCs) Half-Quadrupole-Half-Circle shaped cavity, and confirmed the robustness of the directionality about the shape errors. Based on these, we proposed a hill-climbing algorithm to optimize the ARCs for unidirectional emission. Different evaluation functions of directionality were tested and we suggested using the function of energy contained in a certain angle for highly collimated and unidirectional emission. By this method, we optimized the ARCs to obtain about 0.46 of the total radiated energy in divergence angle of 40º in the far field. This optimization method is very powerful for the shape engineering of ARCs and could be applied in future studies of ARCs with specific emission properties.

## I. INTRODUCTION

ASYMMETRIC symmetric Resonant Cavities (ARCs), which are deformed from the circular cavities, have attracted much attention recently [[1]], [[1]]. Slightly deformed ARCs, in which Whispering Gallery Modes (WGMs) can still exist, possess both high quality (Q) factor and small mode volume. Such advantages make them very potential for applications in nonlinear optics [3], low threshold lasers [4], ultra-sensitive sensors [5], cavity quantum electrodynamics [6] and quantum optomechanics [7]. Furthermore, with the directional emission property, the modes in ARCs can be efficiently excited and collected in free space, regardless of the restriction of near field coupler.

Since the directional emission was observed in experiment for the first time, improving the emission directionality has become a hot topic both theoretically and experimentally. Many boundary shapes have been proposed and demonstrated in experiments for unidirectional emission, such as the Limaçon, Gibbous and the rounded corner triangle cavities [8]-[14]. In previous studies, cavity shapes were specifically defined or randomly chosen with the assistance of computer (see e.g. Ref. [15]). However, there is no evidence that these boundary shapes are optimal for unidirectional emission. Ray simulation has been widely used in these studies, but the reliability of the ray simulation has still not been studied from the view of statistics. Moreover, in practical situations there are always differences between the experimental product and design for ARCs, the influence of which still remains unknown.

In this paper, the basic principles of the unidirectional emission were illustrated by ray dynamics, and the statistical properties of the Monte Carlo simulation of ray dynamics were studied. Moreover, we numerically studied the impact of shape errors induced by the imperfection of fabrication on the directionality, and the robustness of the cavity shape for unidirectional emission was demonstrated. Then, we proposed an optimization algorithm based on ray simulation, and discussed the optimized shapes and corresponding far-field distributions with different criteria of directionality. Finally, unidirectional emission with energy of about 0.46 in angle of 40º in far-field was confirmed by the numerical solution obtained by solving Maxwell equations of an optimal cavity shape [16]-[18].

## II. BASIC PRINCIPLES

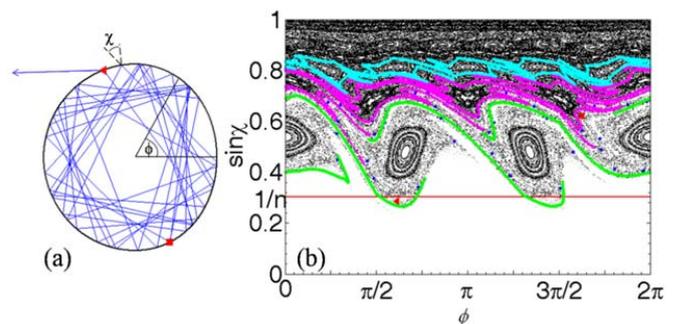

Fig. 1 (a) A typical ray trajectory in real space of light escaping from the Half-Quadrupole-Half-Circle (HQHC) cavity after a sequence of reflections. The deformation of the cavity was characterized with shape parameter $a_2 = 0.11$. (b) Phase space presentation of ray dynamics of the HQHC cavity. The horizontal axis denotes the angle $\phi$ of the reflection point on the boundary; the vertical axis denotes $\sin\chi$, where $\chi$ is the incident angle. Black dots are the mixed Poincaré Surface of Section (SOS) structure of the HQHC cavity including "islands" and "Chaos Sea". Thick blue and red points correspond to the trajectory in (a). The unstable period-3, period-4 and period-5 manifolds are depicted with thick green, magenta and cyan lines, respectively.

Manuscript received October 20, 2012; revised xxxx; accepted xxxx. Date of publication xxxx. This work was supported in part by the National Natural Science Foundation of China under Grant 11204169/11004184, the Foundation of He'nan Educational Committee of China under Grant 2011A140021, the National Fundamental Research Program of China under Grant 2011CB921200, the Knowledge Innovation Project of Chinese Academy of Sciences, and Fundamental Research Funds for the Central Universities.

Fang-Jie Shu is with the Department of Physics, Shangqiu Normal University, Shangqiu 476000, China (e-mail: shufangjie@gmail.com).

Chang-Ling Zou and Fang-Wen Sun are with the Key Laboratory of Quantum Information, University of Science and Technology of China, Hefei, Anhui 230026, China (e-mail: fwsun@ustc.edu.cn).

In the classical ray model, the high-Q WGM is represented as an ensemble of rays, of which reflection angles are greater than the angle of total reflection on the boundary. If the deformation of the boundary from a circle is slight, most rays will maintain large reflection angles in long reflection sequence, giving rise to long lifetime of rays and high intracavity intensity. Only a small portion of the energy radiates out from the curved cavity boundary through barrier tunneling. Meanwhile, some ray trajectories show chaotic motion (Fig. 1(a)), and those rays will refract out after several times of reflection when the incident angle is less than the critical angle of total reflection. Therefore, rays from the islands can escape to the nearby chaos region through dynamical tunneling, and can finally refract out too.

There exist some regions on the boundary where rays tend to refract out, corresponding to the directional emission of ARCs. According to the ray dynamics, Schwefel *et al.* showed that these points and emission directions are sensitive to the shape of the cavity [19]. Thus, the far-field pattern formed by the emission rays can be controlled by designing the boundary shape of the cavity. In order to obtain the cavities which have high-Q and unidirectional emission WGMs, we have proposed a general boundary curve obtained by simple symmetry analysis [15]

$$R(\phi) = \begin{cases} R_0(1 - \sum a_i \cos^i \phi), & \cos\phi \geq 0 \\ R_0(1 - \sum b_i \cos^i \phi), & \cos\phi < 0 \end{cases} \quad (1)$$

where $\phi$ is polar angle, $a_i$ and $b_i$ are deformations of the *i*th order on each side, and $i = 0, 1, 2…$. For instance, the Half-Quadrupole-Half-Circle (HQHC) shape is a kind of shape with all coefficients $a_i = b_i = 0$ except $a_2 \neq 0$ in Eq. (1). It has been demonstrated that in a HQHC microcavity with refractive index n = 3.3, the unidirectional emission is the universal property of high-Q WGMs [15]. The mechanism of unidirectional emission of high-Q modes in HQHC can be explained clearly by the Poincaré Surface of Section (SOS), in which each reflection of rays is represented by Birkhoff coordinates ($\phi$, $\sin\chi$) (Fig. 1(b)). The counter clockwise high-Q rays lay on the upper region of the half-plane with positive $\sin\chi$ in SOS of HQHC. The disperse points in the chaos sea in the upper region can evolve to the lower region through chaos transport, which is accomplished along the manifolds fastened on the unstable fixed points between each island. Rays refract out if their corresponding phase space points are below the critical line (red line in Fig. 1(b)) in the SOS. Moreover, if the segment of a manifold from an unstable fixed point to the critical line is short and steep, more rays will tend to refract out in this shortcut, which means that they will refract from the cavity near the intersection of the manifold and the critical line. As we can see in Fig. 1(b), in the HQHC one branch of manifolds near $\phi = \pi/2$ meets the above criteria. The emission rays from the favorite emission position ($\phi \sim \pi/2$) are roughly tangential to the boundary, forming the peak near $\theta = 180º$ in far-field distribution. Accordingly, clockwise high-Q rays emitting from the lower point of the boundary will strengthen this single main peak in the far-field distribution.

## III. THE MONTE CARLO RAY SIMULATION

Since the wave function of WGMs in ARCs cannot be solved analytically, the precise mode profile and far-field distribution can only be obtained through numerical simulation [16]-[18]. However, the wave simulation is time-consuming. In order to optimize the far-field distribution of ARC, we resort to the Monte Carlo ray simulations. First of all, we need to investigate the repeatability and accuracy of the results of the ray simulations. In the simulation, the initial coordinates ($\phi$, $\sin\chi$) of rays are chosen randomly, which is confirmed as a well approximation for the distribution of high Q WGMs [1]. Then, the sequent reflections and refractions of rays are solved numerically according to specular reflection and Snell's law, respectively. Finally, the far-field distribution is obtained by summing up the leakages of rays, which are determined by modified Fresnel's law [20] where the tunneling on the curved boundary is considered. The amplitude of ray is reduced in each reflection until it reaches the threshold 0.001 of initial amplitude, with the reflectivity calculated by modified Fresnel's law. In addition, the refraction part of the rays are also recorded, and added to get the far field distribution I(θ). Moreover, we only consider the transverse magnetic polarization in the ray simulation, so the corresponding Fresnel's formulas are used.

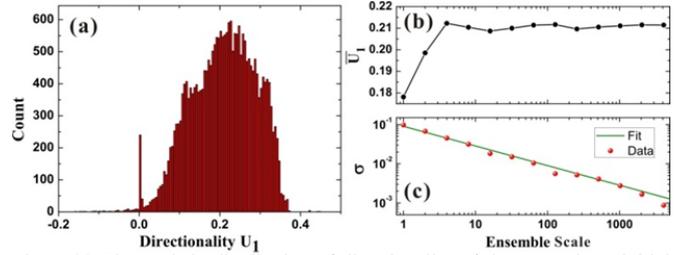

Fig. 2 (a) The statistic distribution of directionality of the rays whose initial positions and reflection angles are in the area where $\sin\chi > 0.6$ through single ray (n = 1) simulation. (b) and (c) are the mean values and the variances of $U_1$ against scale n of ensemble in ray simulation. The number of ensembles is m = 100.

In order to characterize the directionality, we define the directionality function with weighted far-field intensity I(θ) as

$$U_i = -\frac{\int_0^{360} I(\theta)\cos^i\theta\, d\theta}{\int_0^{360} I(\theta)\, d\theta}. \quad (2)$$

Both the amount of energy and some degree of the far-field profile are embodied in this formula, with higher weight of collimation for larger *i*.

For WGMs in ARC, most of the energy is located high above the critical line in SOS, so we will perform ray simulations with initial $\sin\chi > 0.6$, which is very high above the critical line for total reflection $\sin\chi_c = 1/3.3$. First of all, we use $U_1$, which has been utilized in many studies [21], [22], to characterize the general properties of ray dynamics. Obviously, rays starting in different initial conditions will lead to different directionalities. We performed the Monte Carlo simulation of ray dynamics in HQHC and recorded the corresponding $U_1$ in every case of single initial ray. The statistic distribution of $U_1$ is shown in Fig. 2(a). It shows a Gaussian-like distribution, centered at $U_1 \sim 0.21$. There is an abnormal peak around $U_1 = 0$, which is because rays located in the stable island will never leak out with $U_1 = 0$.

To optimize the directionality, we need to get the highly accurate value of directionality of each ARC. Usually, an ensemble of rays distributing randomly in SOS is used to get the directionality. Figs. 2(b) and (c) show the mean values and variances of the $U_1$ for different scales of ensembles. In Fig. 2(c), the variance of $U_1$ of m ensembles (n initial rays are included in each ensemble) reads

$$\sigma = \sqrt{\frac{\sum_{i=1}^{m}(U_{1i} - \overline{U_1})^2}{m}} \quad (3)$$

As the results clearly indicate, when the ensemble scale increases, the mean value of directionality will converge and the variance will reduce. Since the distribution of directionalities for single rays resembles the Gaussian distribution (Fig. 2(a)), their variations are proportion to $1/\sqrt{n}$ (Fig. 2(c)). The larger ensemble corresponds to smaller variation but consumes more computation time. Thus, in the following studies, we choose the ensemble scale of 1000 to ensure the variation $\sigma < 0.005$ and save the computation time.

Actually, the energy of real WGMs is not uniformly distributed in the phase space, thus the directionality obtained by the ray model cannot perfectly correspond to the exact result of wave simulations. The Monte Carlo simulation of single ray indicates that different WGMs correspond to different far-field distributions. Thus, using the ray simulations only the average properties are obtained, which represent the directionality expectation of high-Q WGMs.

## IV. THE ROBUSTNESS OF DIRECTIONAL EMISSION

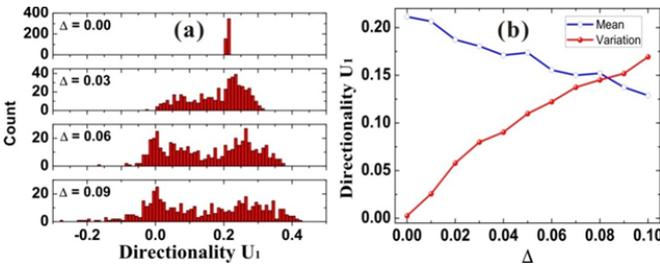

Fig. 3 (a) A statistic of directionalities of 500 ARCs with random shape perturbations on HQHC cavity. $\Delta = 0.00, 0.03, 0.06$, and $0.09$ are the strength of the perturbation . (b) Means and variances of directionality against the perturbation on the boundary. The ensemble scale n = 1000.

Before optimizing the cavity shape, we need to analyze the dependence of directionality on the change of shape parameters, which is another basis of our algorithm of optimization.

To simulate the influence of the shape deformation on the directionality, we added a small random change $\Delta \times (r - 0.5)$ to each shape parameter of HQHC ($a_{2,3,4}, b_{2,3,4}$), where $r \in [0,1]$ is a random number and $\Delta$ is the strength of the perturbation. When $\Delta = 0$ (there is no perturbation), the directionality is about 0.21 (first panel of Fig. 3(a)), with a very small variance caused by finite ensemble scale as discussed in the previous section. When the perturbation strength $\Delta$ increases, the distribution of the directionalities becomes wider, i.e. the variance of the directionalities increases, but the average value of directionalities only changes a little and is still greater than 0.1. The variance against the perturbation is shown in Fig. 3(b), which increases with increasing $\Delta$. Even when $\Delta$ is as large as 0.09, there are still about 40% cavity shapes whose directionalities are better than 0.2.

These results clearly indicate that the HQHC (perhaps other designed unidirectional ARCs as well) is very robust against the experimental fabrication imperfection, and explain the great agreement between the experiment results and the designs in former experimental articles. For example, for an ARC with radius $R = 10\lambda$, the perturbation $\Delta = 0.01$ corresponds to a maximum derivation of boundary $dR = 1.5\Delta R = 0.15\lambda$. The precision of boundary shape error $< 0.15\lambda$ is well within the reach of current lithograph fabrication technique.

On the other hand, the very small variance with small perturbation means that the directionality is a function varying continuously with the change of the shape parameters. Therefore, we can optimize the cavity shape by applying the traditional local search (also called "hill-climbing") optimization process [23].

## V. OPTIMIZATION

Now, we turn to optimize the cavity boundary shape for unidirectional emission. As the general definition of the cavity shape expressed by Eq. (1), 6 parameters determine the shape. Therefore, we can use a 6-element vector $\vec{v} = \{a_2, a_3, a_4, b_2, b_3, b_4\}$ as the optimization object. The algorithm for solving the multi-parameter optimization problem is:

(1) Define the evaluation function of the directionality as $D(\vec{v})$.
(2) Set an initial boundary shape $\vec{v}_0$, and the corresponding directionality $D_0 = D(\vec{v}_0)$ is calculated.
(3) New shape vector is chosen around the old one by random walk method, i.e. $\vec{v} = \vec{v}_0 + \vec{\delta}$, where $\vec{\delta} = \{\delta_i\}$ with $\delta_i = \Delta \times (r_i - 0.5)$, i=1,2…6. The random walk step size $\Delta = 0.01$, and $r_i \in [0,1]$ is a uniform distributed random number.
(4) If $D(\vec{v}) > D_0$, then accept the boundary shape $\vec{v}_0 = \vec{v}$ as a new initial point for optimization with $D_0 = D(\vec{v}_0)$, and repeat steps (3) and (4) till $|D(\vec{v}) - D_0| < \varepsilon$.

From the analysis of statistical property and robustness of directionality, the ray simulation can give an accurate evaluation of directionality $D(\vec{v})$, and then the random walk with small step size will result in a slow "climbing up" of $D(\vec{v})$ approaching to a local maximum. A restriction of parameters, $R(\theta) \in [0.5, 1.5]$ for arbitrary $\theta$, is applied during the optimization process to avoid non-physical cavity shapes.

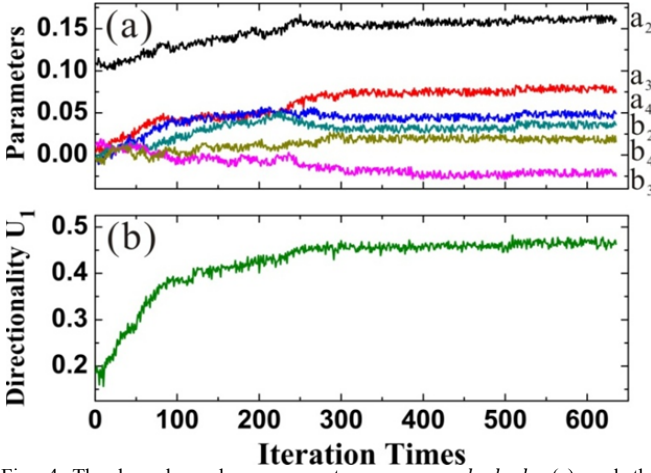

Fig. 4 The boundary shape parameters: $a_2, a_3, a_4, b_2, b_3, b_4$ (a) and the directionality $U_1$ (b) evolve against the iteration times in the optimization process, with the start point $\{a_2 = 0.11, a_3 = 0, a_4 = 0, b_2 = 0, b_3 = 0, b_4 = 0\}$ and $U_1 = 0.21$.

Starting from the HQHC shape of $\vec{v}_0 = \{0.11, 0, 0, 0, 0, 0\}$, and using the directionality $U_1$ as evaluation function, we obtained the optimized cavity shape and U by the optimization algorithm. Shape parameters and $U_1$ of the intermediate shape against the iteration times during the optimization process are shown in Fig. 4. The parameters show random fluctuations which reflect the random walk of shape. The $U_1$ obviously increases with increasing iteration times, and finally reaches a saturation value of about 0.47. It only takes about 300 iterations for the $U_1$ to increase from 0.20 to a more than doubled value of about 0.47, showing that the optimization algorithm is very effective.

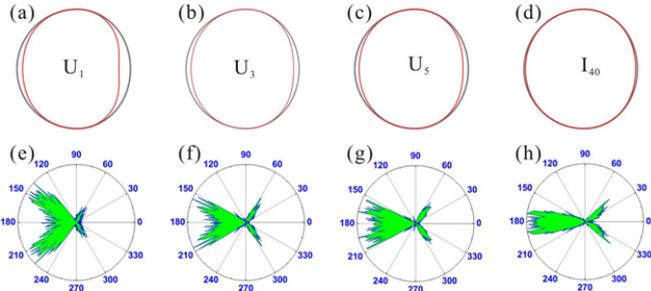

Fig. 5 The optimized boundary shapes (red lines in (a), (b), (c), and (d)) obtained by the same optimization process but different evaluation functions ($U_1$, $U_3$, $U_5$ and $I_{40}$) and the corresponding far-field distributions ((e), (f), (g), and (h)). The corresponding shape parameters are listed in table I. Initial shape is HQHC with $a_2 = 0.11$.

As shown in Fig. 5(a), the ARC boundary shape (red line) optimized by setting $D(\vec{v}) = U_1$ is just slightly deformed from the original HQHC cavity (black line). Though most energy emits to the left as expectation, the far-field distribution shows two peaks around $\theta = 150º$ and $\theta = 210º$ resulting in a very large divergence angle (about 110º in far-field, see Fig. 5(e)). This broad emission angle of the optimized cavity is because that the weight function cosθ in the evaluation function $U_1$ is not sensitive to divergence.

TABLE I
PARAMETERS OF OPTIMIZED CAVITIES

| Shape | Parameters | | | | | |
|---|---|---|---|---|---|---|
| | $a_2$ | $a_3$ | $a_4$ | $b_2$ | $b_3$ | $b_4$ |
| $U_1$ | 0.178 | 0.040 | 0.084 | 0.020 | -0.029 | 0.045 |
| $U_3$ | 0.124 | 0.046 | 0.031 | 0.035 | -0.017 | 0.032 |
| $U_5$ | 0.136 | 0.008 | 0.054 | 0.007 | -0.019 | 0.046 |
| $I_{40}$ | 0.118 | 0.003 | 0.020 | 0.005 | 0.008 | 0.028 |

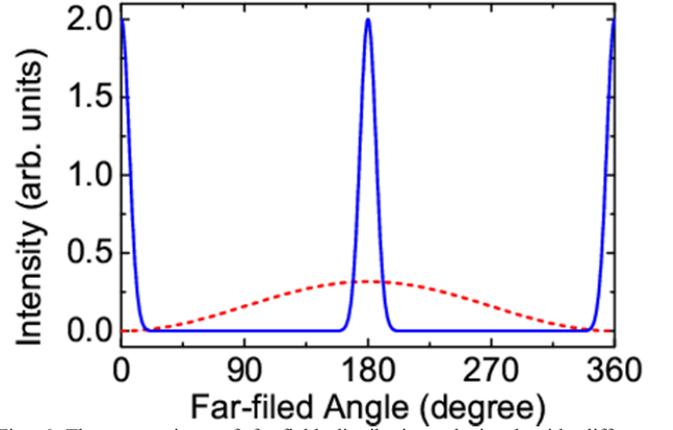

Fig. 6 The comparison of far-field distributions depicted with different functions $I(\theta) = 2\cos^{100}(\theta)$ (blue solid line) and $I(\theta) = \sin^2(\theta/2)/\pi$ (red dashed line). The former one shows a bidirectional emission distribution with narrow divergence angle, while the later one hardly shows a unidirectional emission distribution.

To get a better divergence angle of unidirectional emission, we use the $U_3$ and $U_5$ as evaluation functions of optimization. Their weight functions are more concentrated near $\theta$ of 180º. As shown in Fig. 5 (b), (c), (f) and (g), the results indicate that better collimated emission could be obtained by applying $U_i$ with higher order index i. However, the divergence of optimized boundary shape by $U_5$ is still wider than 60º. This is because that the unidirectional emission is not strictly dominated by higher order $U_i$. For example, there are two far-field distributions described by $I(\theta) = 2\cos^{100}(\theta)$ (blue line) and $I(\theta) = \sin^2(\theta/2)/\pi$ (red dashed line) respectively (Fig. 6), where the intensities are normalized to $\int_0^{2\pi} I(\theta)d\theta = 1$. For the two-peaks (bidirectional emission) far-field distribution (blue solid line), we have $U_1 = U_3 = U_5 = 0$. For the single-peak (unidirectional emission) with larger divergence (red dashed line), we have $U_1 = 0.25$, $U_3 = 0.1875$, and $U_5 = 0.15625$. So, which one of the distributions has better directionality? In terms of the U function, red dashed line has a much better directionality. However, from the view of the real space distribution, the blue line shows that about 50% of total radiation energy is contained in an angle of 30º, while there is only 16% in the red dashed line case.

Actually, the efficient collection and excitation of WGMs in ARC in experiments require the directional emission to contain as much energy as possible in a given collection angle. Considering the actual requirement, we define the directionality of unidirectional emission as

$$I_{\theta_d} = -\frac{\int_0^{360} I(\theta) g(\theta) d\theta}{\int_0^{360} I(\theta) d\theta} \quad (4)$$

with the weight function

$$g_{\theta_d}(\theta) = \begin{cases} 1 \ldots \theta \in (180 - \frac{\theta_d}{2}, 180 + \frac{\theta_d}{2}) \\ 0 \ldots \text{else} \end{cases} \quad (5)$$

which just concerns the energy in divergence of $\theta_d$. For instance, using the $I_{40}$ ($\theta_d = 40°$) as the high collimation evaluation function, we obtain the optimized boundary (Fig. 5(d)) with the parameters given in Table I, and the far-field distribution (Fig. 5(h)) which shows a significant directionality.

TABLE II
DIRECTIONALITIES OF OPTIMIZED CAVITIES

| Cavity shape optimized by | Directionality | | | |
|---|---|---|---|---|
| | $U_1$ | $U_3$ | $U_5$ | $I_{40}$ |
| $U_1$ | 0.476759 | 0.379011 | 0.307480 | 0.200776 |
| $U_3$ | 0.393899 | 0.409686 | 0.390210 | 0.345904 |
| $U_5$ | 0.382273 | 0.411986 | 0.404886 | 0.399466 |
| $I_{40}$ | 0.246787 | 0.281854 | 0.299091 | 0.458326 |

In Table II, we compare the directionalities evaluated by different functions for the optimized boundaries with each other. Each boundary shape attains the highest value of directionality assessed by the corresponding evaluation function. The only exception that the value of $U_3$ under optimized $U_5$ shape is a little higher than that of $U_5$ under $U_5$ itself, is attributed to the small variance of the directionality of ray simulation. The $I_{40}$ shape has the most energy of about 0.46 in divergence of 40°, although its $U_1$, $U_2$, and $U_3$ are obviously smaller than other shapes.

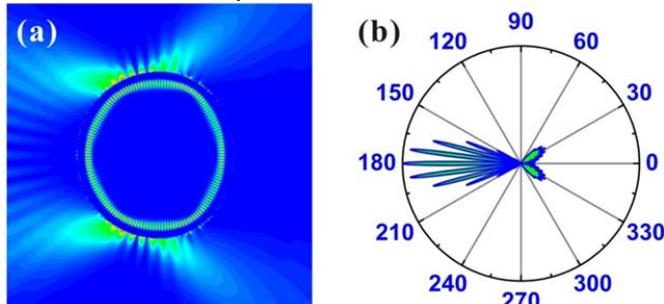

Fig. 7 The near field (a) and far-field (b) distributions of $I_{40}$ optimized cavity calculated by the wave simulation, with $kR = 25.44$ and $Q = 9.82 \times 10^7$.

Finally, we checked the $I_{40}$ optimized unidirectional emission cavity (Figs. 5(d)) by the boundary element method [18-19]. The quasi-fundamental WGM with the transverse magnetic polarization is shown in Fig. 7, with $kR = 25.44$ and high-Q factor $9.82 \times 10^7$ [20]. The far-field distribution shows good unidirectional emission, which greatly consists with the ray simulation results (Fig. 5(h)). We calculated $U_1 = 0.223$ and $I_{40} = 0.437$ for the far field distribution in Fig. 7(b), which are also consistent with our ray prediction perfectly (Table II).

## VI. CONCLUSION

In conclusion, we proposed and demonstrated the optimization of the ARC boundary shape for unidirectional emission, with the cavity refractive index n = 3.3. The statistical properties of the ray dynamics were studied in detail. Based on the statistical properties of the ray dynamics and the directionality against the perturbation of the boundary shape, the algorithm of local searching was given. Several different evaluation functions of directionality were considered in the optimization, and the results showed that the evaluation function could affect the final optimal shape greatly. We proposed a function $I_{40}$ to evaluate the directionality for the requirement of highly efficient collection in experiment, and the unidirectional emission was confirmed by the wave simulation. We believe that the optimization method demonstrated here can also be used for other aims of boundary engineering.

**Fang-Jie Shu** received the B. S. and Ph.D. degrees in physics from USTC in 2002 and 2007, respectively. He is now an Associate Professor at Shangqiu Normal University, He'nan, China.

**Chang-Ling Zou** received the B.S. degree from the University of Science and Technology of China (USTC), Hefei, China, in 2008. He is currently pursuing the Ph.D. degree in optics at USTC.

**Fang-Wen Sun** received the B.S. and Ph.D. degrees from USTC in 2001 and 2007, respectively. He is now an Associate Professor at USTC.